# A Class of Exactly Solvable Hamiltonians for S=1/2 Quantum Magnets with Spinless Fermionic Excitations in Higher Dimensions

Sumiran Pujari[1, a)]

[1]Department of Physics, Indian Institute of Technology Bombay, Powai, MH 400076, Mumbai
[a)]sumiran.pujari@iitb.ac.in

**Abstract.** This contribution summarizes the main results of a work on exactly solvable Hamiltonians for quantum magnets. A class of Hamiltonians which supports fractionalized spinless fermionic excitations in dimensions greater than one is written down. A well-known one-dimensional example is that of S=1/2 spin chains with Luttinger liquid physics and spinless fermionic excitations that are also called spinons. A well-known two-dimensional example is that of Kitaev's S=1/2 honeycomb model with bond-dependent magnetic couplings which supports Majorana fermionic excitations. The class of models to be discussed here also exploits bond-dependent couplings in a different way to non-perturbatively stabilize spinless fermionic spinons and also Majorana fermions. A detailed account of these results including all the supporting technical information is being prepared for publication elsewhere. Funding support from SERB-DST, India via Grants No.MTR/2022/000386 and partial support by Grant No. CRG/2021/003024 is also acknowledged.

## BRIEF INTRODUCTION

Exactly solvable models provide cornerstones in condensed matter physics with strong correlations. They provide representatives of some phase of condensed matter or perhaps even a quantum phase transition [1] that can be studied in much greater detail than generic points in the phase diagram. Here we concern with models that have spinless fermionic excitations in two and higher dimensions. The solution method relies on the fact that the models are designed to have an extensive number of conserved quantities such that the full eigenspectrum can be obtained. This is similar in spirit to the Kitaev model [2] or even more simply to the classical Ising model. Our method mixes this general approach with the Jordan-Wigner transformation [3] that is well-known for facilitating the solution of the quantum Ising model [4,5]. We are not attempting to obtain the free energy from the solved eigenspectrum. Correlations can be obtained from the eigenfunctions [6], but that is not the focus here. Also in common with the Kitaev model, these models are built out of bond-dependent magnetic couplings that are not spin rotation symmetric. However the nature of the conserved quantities in these models is quite different than the Kitaev model.

Spinless fermionic excitations in two dimensions are also expected in spin rotation symmetric Heisenberg magnets on frustrated geometries. These models are almost never exactly solvable though and generally require approximate mean-field methods or semi-rigorous field theoretic arguments to argue for the existence of the spinon excitations [7]. The two-dimensional Shastry-Sutherland model [9] and the one-dimensional Ghosh-Majumdar model [10] are perhaps the only two well-known examples with Heisenberg couplings that support gapped valence bond solids whose triplet excitations can be considered as a bound state of two confined spinons. This is not as satisfactory from the point of view of arguing for the existence of spinon excitations. The class of bond-dependent models reported here thus gives a non-perturbative proof-of-principle of spinon excitations in higher dimensions. These models can be tweaked in a straightforward manner to support Majorana excitations as well, which would then form cousins of the Kitaev model.

In the rest of the paper, we will discuss the basic idea behind the construction of this class of models and the resultant conserved quantities. Then we discuss the array of ground states and fermionic excitations that result from this construction on different kinds of lattices. We will touch upon the stability of these states upon deforming the Hamiltonians away from the exactly solvable constructions. Some remarks on material possibilities and relevance to quantum engineering will also be mentioned at the end.

## RESULTS

The models are constructed as appropriate higher dimensional grids of XY spin chains and ZZ Ising spin chains. We choose Z-axis as the spin quantization axis for this work. Some possibilities with or without triangular motifs in two dimensions are illustrated in Fig. 1. As shown in this figure, these models have XY spin chains separated from each other by intervening spins. These intervening spins are coupled via ZZ Ising couplings to the rest of the system. The spins on the XY spin chains will be referred to as "on-chain" spins, whereas the intervening spins will be referred to as "off-chain" spins. One may replace the XY couplings of the on-chain spins with asymmetric XY couplings all the way to the XX Ising limit without losing solvability. Note this solvable deformation in the Ising limit implies that the on-chain (XX) Ising couplings and the off-chain (ZZ) Ising couplings being along orthogonal directions in spin space somewhat similar to the Kitaev honeycomb model. Furthermore, the intervening off-chain spins can be longer ZZ Ising chains as well while retaining solvability but we are not pursuing that here. Similar constructions can be straightforwardly made in higher dimensions also.

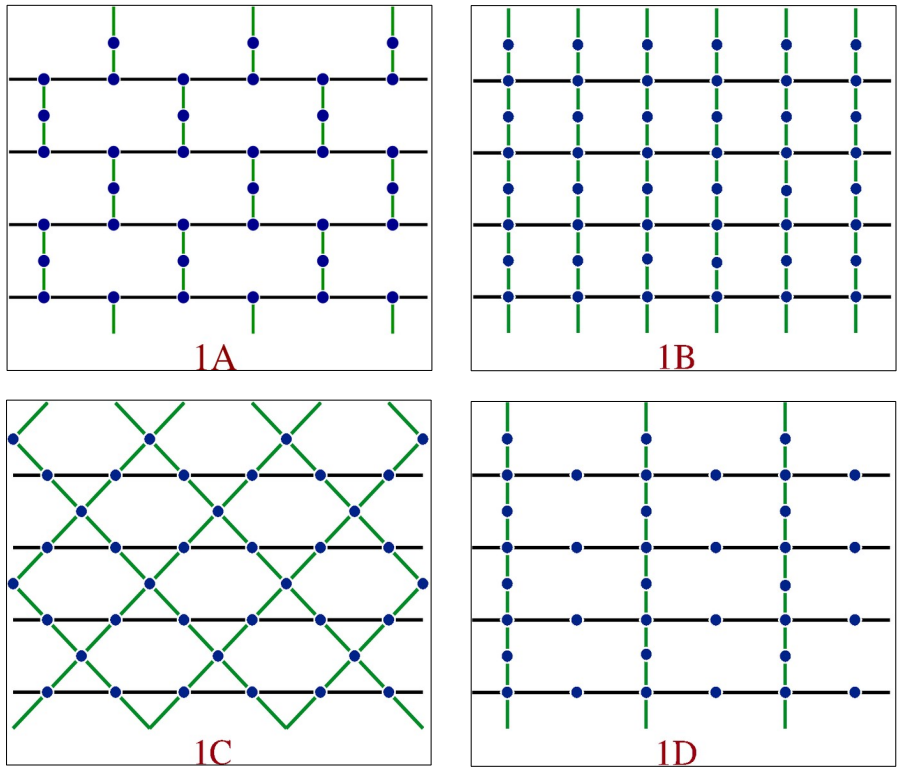

**FIGURE 1.** This figure shows four models in the exactly solvable class of S=1/2 quantum spin Hamilonians put forth in this work. The blue dots represents the quantum S=1/2 degrees of freedom. The black solid lines represents XY couplings between these degrees of freedom. These couplings may be deformed to anisotropic XY couplings all the way to XX Ising limit without losing Jordan-Wigner solvability. The green solid lines represents ZZ Ising couplings between these degrees of freedom.

The solvability rests on the fact that the off-chain spins are conserved quantities, i.e. their Z-operators commute with the Hamiltonian by virtue of the Ising nature of their couplings to the rest of the system. Thus, one may fix their Z-values and diagonalize the on-chain Hamiltonians. The fixed off-chain spins provide a background field for the on-chain spins. Such a situation can be solved by the Jordan-Wigner transformation which leads to a spinless fermionic spectrum. Furthermore, the on-chain spin chains can be considered independently from each other. One can solve for the spectrum of just one of them and transfer the results to the rest. This would give the spectrum of the full two or higher dimensional system. The solvability is also preserved in presence of an external field along the quantization axis for the on-chain spins. Here we are assuming that the number operator of Jordan-Wigner fermions corresponds to spin polarization along the quantization axis as is often done.

One technical point of note is that the different configurations of the conserved off-chain spins correspond to different sectors or blocks of the Hamiltonian. One thus needs to find that off-chain spin configuration which leads to the lowest energy in order to find the ground state. This is analogous to finding the conserved plaquette flux sector that leads to the lowest energy in the Kitaev model [2,11]. In our models the conserved quantities are site-local, which are in contrast to that of the Kitaev model where the conserved quantities are plaquette-local and comparatively more intricate.

Below we summarize the results obtained for several cases:

1. For the first model shown in Fig. 1A on a brick-work like lattice without any triangular motifs, we find that the ground state sector obtains for an antiferromagnetically ordered configuration for the conserved off-chain spins. The corresponding on-chain many-body ground state is a gapped half-filled spinless fermion sea. The underlying spinless fermion band structure comprises two bands with a gap between them that corresponds to nearest neighbour hopping fermions with the value of the chemical potential alternating from site to site. The gap can be understood as being "induced" by the background field set up by the antiferromagnetic configuration of the off-chain spins. The word "induced" is in quotes because the off-chain configuration is not fixed beforehand but is rather dynamically selected due to energetics of this model, even though the off-chain spins are not involved in the quantum dynamics at the level of the microscopic Hamiltonian. The ground state is thus like a Luttinger liquid with a gap that has opened up at the Fermi level. This also leads to a staggered magnetization for the on-chain spins as well. This state can thus be called as a spinless fermion solid possessing gapped spinon excitations.

2. The second model shown in Fig. 1B is a variation on Fig. 1A and has similar properties.

3. For the third model shown in Fig. 1C on a kagome lattice with triangular motifs, we find that the ground sector rather obtains for a ferromagnetically ordered configuration for the conserved off-chain spins. The corresponding on-chain many-body ground state is a gapless partially filled spinless fermion sea. The underlying spinless fermion band structure comprises a single band with a dispersion that corresponds to nearest neighbour hopping fermions. The partial filling can be understood due to the chemical potential shift induced by the background field set up by the ferromagnetic configuration of the off-chain spins. The word "induced" is again in the sense previously used. The ground state therefore is like a stack of Luttinger liquids with gapless spinon excitation along the on-chain direction. This will thus lead to strongly anisotropic spin transport along the on-chain direction.

4. Presence or absence of triangular motifs has a great bearing on the ground state physics as seen in the previous points. It may be remarked here that the exact sign structure of the XY and ZZ couplings in the microscopic Hamiltonians is not relevant. In fact, there are simple unitary transformations that can give other signs of these couplings and thereby the resultant ground states.

5. We find that there is an appreciable energy gap for changing the configuration of the off-chain spins. This implies that the ground state physics described above is stable to such fluctuations. A different way to put this is as follows: the conserved nature of the off-chain spins can be thought of as absolutely localized spinless fermions under the Jordan-Wigner transformation. Then the equivalent statement is that there is a gap to exciting these localized spinless modes away from their ground state configurations, i.e. antiferromagnetic off-chain configuration for models 1A and 1B, and ferromagnetic off-chain configuration for model 1C.

6. The previous point also has a bearing on the stability of these states to deformations of the Hamiltonian away from the solvable constructions, in particular to ZZ couplings for on-chain bonds and XY couplings for off-chain bonds. Having an energy gap for changing the ordered configuration of the off-chain spins also implies stability with respect to XY deformations for off-chain bond couplings. This is because such off-chain XY deformations are precisely those quantum processes that can drive quantum fluctuations in the occupation number of off-chain spins which is conserved in the exactly solvable limit by the design of our construction, i.e. when the off-chain XY deformations are absent. Having a gap to these fluctuations implies that if the energy scale for these off-chain XY deformations is smaller than the gap mentioned in the previous point, then the ground state will be stable to such deformations. Stability with respect to on-chain ZZ Ising deformations is easier to argue. It follows from the standard arguments for the stability of the Luttinger liquid physics in one-dimensional spin chains [12]. This implies that the spinless fermionic excitations should be stable to both kind of deformations that are to be naturally expected in a realistic material.

7. As remarked before, the solvability is preserved if one moves away from XY couplings to anisotropic XY couplings all the way to XX Ising limit for the on-chain spins. In this Ising limit, we find that for model 1C, the ground state is again a ferromagnetic configuration for the off-chain spins, with an XX Ising ordered/disordered ground state with gapped Majorana excitations for the on-chain spins. In terms of Jordan-Wigner fermions, this is a non-trivial/trivial superconducting ground state. This is very analogous to the physics of the quantum Ising model with the field value being set by the off-chain ZZ Ising couplings. For model 1A, something more interesting is seen. The ground state is actually *any* configuration for the off-chain spins, with an XX Ising ground state with gapped Majorana excitations for the on-chain spins. In other words, all sectors defined by the off-chain spin configurations have the same lowest energy state. The many-body ground state is thus a coexistence of an Ising magnet for the on-chain spins and a completely paramagnetic state for the off-chain spins at zero temperature. This extensive degeneracy in the off-chain spins can be understood as additional conserved quantities that anti-commute with the conserved off-chain spins (and commute with the Hamiltonian). These additional conserved quantities are products of X-operators on the (finite) ZZ-Ising segments of the lattice. For model 1A, they are thus the product of 3 X-operators for each ZZ-Ising segment shown as green solid lines in Fig. 1. In a realistic situation, there will be other (smaller) couplings which will then select the "true" ground state configuration for the off-chain spins to accord with the third law of thermodynamics. A natural expectation in such a situation is to presume the existence of a (possibly gapless) "slow" mode primarily composed of the off-chain spins as excitations on top of the ground state, apart from the gapped Majorana excitations from the on-chain spins which will be higher in energy compared to the presumed slow mode. The solvable limit can be said to have an extensive number of zero modes from the perspective of this expectation. The absence of triangular motifs is not an absolute requirement for this physics; rather one only requires finite ZZ-Ising segments with at least one conserved spin in the spirit of the above constructions connected through solely XX-Ising (or solely YY-Ising) segments.

8. A bond-dependent Hamiltonian along the above lines can also be written down on the Lieb lattice as shown in Fig. 1D. In this case, there are additional on-chain spins that are also conserved quantities. This actually leads to a classical ordered ground state without any spin liquid physics.

## CONCLUDING REMARKS

We conclude with several remarks on the above results:

- The fermionic excitations of this class of models have a stripy character as opposed to the Majorana excitations in the Kitaev model which have full two-dimensional character. The conserved quantities are site-local as mentioned before and thus are simpler than the plaquette-local conserved quantities of the Kitaev model.

- The stripy character of the excitations might be useful from a quantum engineering perspective. The on-chain spins can be thought of as highways and the off-chain spins as guardrails for quantum information traffic. The lattice structures motivated by this class of models can potentially form the basis of an architecture for qubits for quantum information processing.

- Material possibilities of these models do not seem unrealistic. One needs an insulating material such as oxides which has an appropriate grid of easy-axis and easy-plane spin chains. It does not necessarily need spin-orbit coupling physics that are crucial to realize Kitaev physics [13]. However, bond-dependent magnetic couplings need anisotropies in spin space that perhaps demands specific crystal structures as the matrix for the magnetic ions hosting the S=1/2 degrees of freedom. This could be an interesting avenue of exploration for the materials design community. The strongly anisotropic spin transport expected in these models can then potentially be realized in the laboratory and may be relevant for quantum or classical technologies.

- Finally, from a theoretical quantum magnetism perspective, these models lead to interesting states. On lattices without triangular motifs, they realize an antiferromagnetically ordered ground state with gapped spinless fermionic excitations or spinons. This spinless fermionic solid can be contrasted with conventional antiferromagnets with gapless bosonic spin wave excitations. On lattices with triangular motifs, they realize a ferromagnetically ordered ground state with gapless spinons again in contrast to conventional ferromagnets with bosonic spin wave excitations. On replacing the XY couplings of on-chain spins with XX Ising couplings, we also find a very unusual state in absence of triangular motifs wherein an Ising magnet with Majorana excitations composed from the on-chain spins coexists with a paramagnetic state composed from the off-chain spins in the ground state. This state will lead to a two-step entropy release similar to in the Kitaev model in presence of thermal fluctuations with the first step corresponding to the entropy release due to the off-chain spins and the second step corresponding to the on-chain spins. The relative magnitude of the two steps will depend on the ratio of the number of off-chain spins to on-chain spins. This ratio is 1:2 for model 1A and 1:1 for model 1B. Thus, these states can be considered as simpler variants of the Kitaev spin liquid.